\documentclass[aps, prb,twoside,twocolumn,english, floatfix, superscriptaddress]{revtex4-1}

\usepackage{textgreek}
\usepackage[T1]{fontenc}
\usepackage[utf8]{inputenc}
\usepackage{xcolor}
\usepackage{units}
\usepackage{babel}
\usepackage{textcomp}
\usepackage{amsmath}
\usepackage{graphicx}
\usepackage[colorinlistoftodos]{todonotes}

\begin{document}

\preprint{APS/123-QED}

\title{Inelastic spin-wave beam scattering
  by edge-localized spin waves in ferromagnetic thin film}% Force line breaks with \\

\author{Pawel Gruszecki}
\email[corresponding author:~]{gruszecki@amu.edu.pl}
\affiliation{Institute of Spintronic and Quantum Information, Faculty of Physics, Adam Mickiewicz University, 61-614 Poznan, Poland}

\author{Konstantin Y. Guslienko}
\affiliation{Division de Fisica de Materiales, Depto. Polimeros y Materiales Avanzados: Fisica,
Quimica y Tecnologia, Universidad del Pais Vasco, UPV/EHU, 20018 San Sebastian, Spain}
\affiliation{IKERBASQUE, the Basque Foundation for Science, 48009 Bilbao, Spain}

\author{Igor L. Lyubchanskii}
\affiliation{Donetsk Institute for Physics and Engineering named after O. O. Galkin
(branch in Kharkiv) of the National Academy of Sciences of Ukraine,
03028 Kyiv, Ukraine}
\affiliation{Faculty of Physics, V. N. Karazin Kharkiv National University, 61022 Kharkiv, Ukraine}

\author{Maciej Krawczyk}
\affiliation{Institute of Spintronic and Quantum Information, Faculty of Physics, Adam Mickiewicz University, 61-614 Poznan, Poland}

\date{\today}% It is always \today, today,

\begin{abstract}
 %Spin waves at certain conditions can be confined in particular regions of the sample, for example in a potential well created by the demagnetizing field near the film’s edge. Typical frequencies of the edge-localized spin waves lay below the bottom of the film spin-wave spectrum. 
 
 Spin waves  are promising chargeless information carriers for the future, energetically efficient beyond-CMOS systems.  Among many advantages there are the ease of achieving nonlinearity, the variety of possible interactions, and excitation types. 
 Although the rapidly developing magnonic research has already yielded impressive realizations, multi-mode nonlinear effects, particularly with the propagating waves and their nanoscale realizations, are still an open research problem.
 %While rapidly developed magnonic research  already yielded impressive results, developing inter-modal nonlinear effects for propagating modes and scaleable is an open research problem.
  We study theoretically the dynamic interactions of the spin waves confined to the edge of a thin ferromagnetic film with the spin-wave beam incident at this edge. We found the inelastically scattered spin-wave beams at frequencies increased and decreased by the frequency of the edge spin-wave relative to the specularly reflected beam. 
  %This three-magnon scattering phenomenon is the  magnonic counterpart of the Brillouin light scattering by spin waves. 
  We observed a strong dependence of the angular shift of the inelastic scattered spin-wave beam on the edge-mode frequency, which allowed us to propose a magnonic demultiplexing of the signal encoded in spin waves propagating along the edge. Since dynamic magnetostatic interactions, which are ubiquitous in the spin-wave dynamics, are decisive in this process, this indicates the possibility of implementing the presented effects, also in other configurations and their use in magnonic systems.
  
  %We show that dynamic magnetostatic interactions are decisive in this process, which indicates its ubiquity and the possibility of its implementation in various other configurations. Moreover, we observed an angular shifts between the primary incident  and the frequency dependent  reflected spin-wave beams, which allows us to propose magnonic demultiplexing of the signal coded in the edge spin waves. 

\end{abstract}

\maketitle

%\tableofcontents

\section{Introduction}\label{Sec:introduction}

Nonlinear interactions are ubiquitous in nature and are indispensable in information processing devices, regardless of the carrier used for coding. The nonlinearity of spin-wave (SW) dynamics has long attracted the attention of researchers, and the topic is still very relevant in the emerging field of magnonics; see Refs.~[\onlinecite{wigen1994, cottam1994, gurevich1996,Nikitov2015,Bracher20171, hula2021spin, pirro2021advances}] and  references therein. Among the many nonlinear effects that have been  examined for  application in magnonics is the decrease in the static magnetization component with increasing excitation power and the correlated change in  SW frequency.\cite{qiwang2020nonlinear} The process of parametric pumping has been explored as a promising tool for the amplification of SWs, but it still awaits a suitable method for its  integration, a necessary step in the construction of magnonic circuits. There are also other multiple-wave types of interaction between SWs, involving waves of the same origin, such as surface magnetostatic SWs~\cite{boardman1988, kazakov1997} and dipole-exchange SWs,\cite{pereira2003} allowing for  generation of  second and third harmonics. A number of studies have been devoted to the investigation of three-  and four-magnon collision processes, with different aspects of the nonlinear SW interactions being considered  and various possible applications  proposed.\cite{boardman1988, demidov2008, zhitomirsky2013, liu2019, verba2019, slobodianiuk2019nonlinear} These include, confluence of two waves into one or splitting one wave into two waves,\cite{Camley2014,Ordonez2009} cascading processes resulting in frequency change and nonlinear damping,\cite{costa2000,pereira2003}, generation of the ultrashort spin-wave pulses,\cite{synogach2000} self-focusing and  collapse,\cite{bauer1997direct}  frequency combs generation,\cite{Khivintsev2011} and application in spintronic \cite{kurebayashi2011} and signal processing devices \cite{melkov2009,sadovnikov2017nonlinearMagnonics}.
There have also been other spectacular achievements in this area, such as Bose--Einstein condensation of  magnons at room temperature and the demonstration of a pure magnon transistor.\cite{Demokritov2006,chumak2014magnon}

Interactions between waves of different nature allow for  transfer and control of dynamics in distinct subsystems, such as chiral emission of SWs by a gyrotropic mode of skyrmion \cite{chen2021chiral} or interaction between propagating bulk SWs (B-SWs) and domain-wall oscillations.
%For instance, a chiral emission of SWs by a gyrotropic mode of skyrmion was demonstrated.\cite{chen2021chiral}
Domain walls are considered as channels for SWs, suitable for guiding waves even along curved and very narrow paths, and therefore, they have  attracted much attention,\cite{YU2021} also regarding nonlinear effects.
In particular, the three-magnon interaction in a thin ferromagnetic film between an incident SW of frequency $f_0$ and the domain-wall flexure oscillations at  frequency $\nu$ where investigated \cite{dadoenkova2019}. As a result of this interaction, the SWs are scattered at frequencies $f_0+\nu$  and $f_0-\nu$, thus this phenomenon can be considered as  an inelastic scattering of SWs (ISSW) on domain-wall oscillations. This effect is analogous to the Brillouin light scattering (BLS) by SWs,\cite{cottam1986light, liu2019} with  the incident and scattered electromagnetic waves replaced by SWs, and the SWs replaced by domain-wall \cite{dadoenkova2019, zhou2021spin, zhang2018} or skyrmion\cite{wang2021magnonic} oscillations.  In Ref.~[\onlinecite{zhang2018}], a  three-magnon process was also proposed  as a tool to infer the information guided by SWs along the domain wall. Wang \emph{et al.}\cite{wang2018probing} proposed to use a three-magnon interaction between incident plane B-SWs and an SW confined in a region with  Dzyaloshinskii-Moriya interaction as a tool to determine the strength of the Dzyaloshinskii-Moriya interaction.

From the points of view of both  fundamental physics and  applications,  SWs located at  edges or interfaces in planar structures are of particular interest. The edge confinement can be due to topological protection, wave properties, or a nonuniformity of the internal fields.\cite{Shindou2013,Lisenkov2016,Lara2013,Guo2013}
In a finite structure with a magnetization component normal to the film edge,  surface magnetic charges create a static demagnetizing field, assuming a not ellipsoidal sample. In that case, the demagnetizing field is nonuniform and forms wells in the internal magnetic field that is suitable for confining low-frequency SW oscillations, called edge SWs (E-SWs).\cite{Jorzik2002,Kruglyak2006} In the case of  a thin semi-infinite film or a waveguide with an in-plane external magnetic field directed perpendicularly to the edge, the demagnetizing field creates  wells, enabling  guidance of the SWs  along the edge.\cite{Bailleul2003,gruszecki2014goos,gruszecki2015influence}. 
However,  neither the nonlinear interactions of B-SWs with E-SW beams nor the exchange of energy and information between these two types has  yet been investigated.

In the paper, we investigate the inelastic scattering of B-SWs by E-SWs in a thin ferromagnetic film through an analytical  description and micromagnetic simulations.  We analyze the interaction of an obliquely incident B-SW beam (frequency $f_0$) with the E-SWs (frequency $\nu$) and find  secondary B-SW beams at frequencies $f_0-\nu$ and  $f_0+\nu$. We show that dipolar interactions are responsible for the nonlinearity related to the three-magnon process. The described effects are promising for the extension to other types of E-SWs.
Interestingly, we observe that the angular shifts between the elastically reflected B-SW beam and the inelastically scattered B-SW beams are dependent on frequency. A clear explanation of this effect is obtained on the basis of energy and momentum conservation laws supplemented with the  isofrequency contour analysis.
Subsequently, we propose an application of   inelastic three-magnon scattering to demultiplex information carried by E-SWs at different frequencies. 
Finally, we analyze the efficiency of B-SW scattering at the edge SWs and show that the efficiency of the assisted splitting (also known as stimulated splitting) \cite{wang2018probing, zhang2018} is significantly greater than the confluence process.
Because the proposed effect shall exist in a basic thin-film geometry, it is feasible for experimental demonstration and can have various applications in magnonics, such as in  nonlinear magnonic nano-ring resonators, \cite{qiwang2020nonlinear} nonlinear SW interferometers for power-selective suppression of pulsed microwave signals,\cite{ustinov2007ferrite} nonlinear SW logic gates, \cite{ustinov2019nonlinear} or  omnidirectional SW array antennas. \cite{song2019omnidirectional}

The paper is organized as follows. In Sec.~\ref{Sec:problemFormulation}, we describe the system under investigation. In Sec.~\ref{Sec:model} we introduce the analytical description of inelastic scattering of SWs. Then, in Sec.~\ref{Sec:results}, we present the results of micromagnetic simulations and analyze the efficiency of B-SW scattering. Finally, we summarize our results in Sec.~\ref{Sec:conclusions}.
The appendix with details of micromagnetic simulations is placed at the end of the paper.

\begin{figure}[t!]
\includegraphics[width=8.6cm]{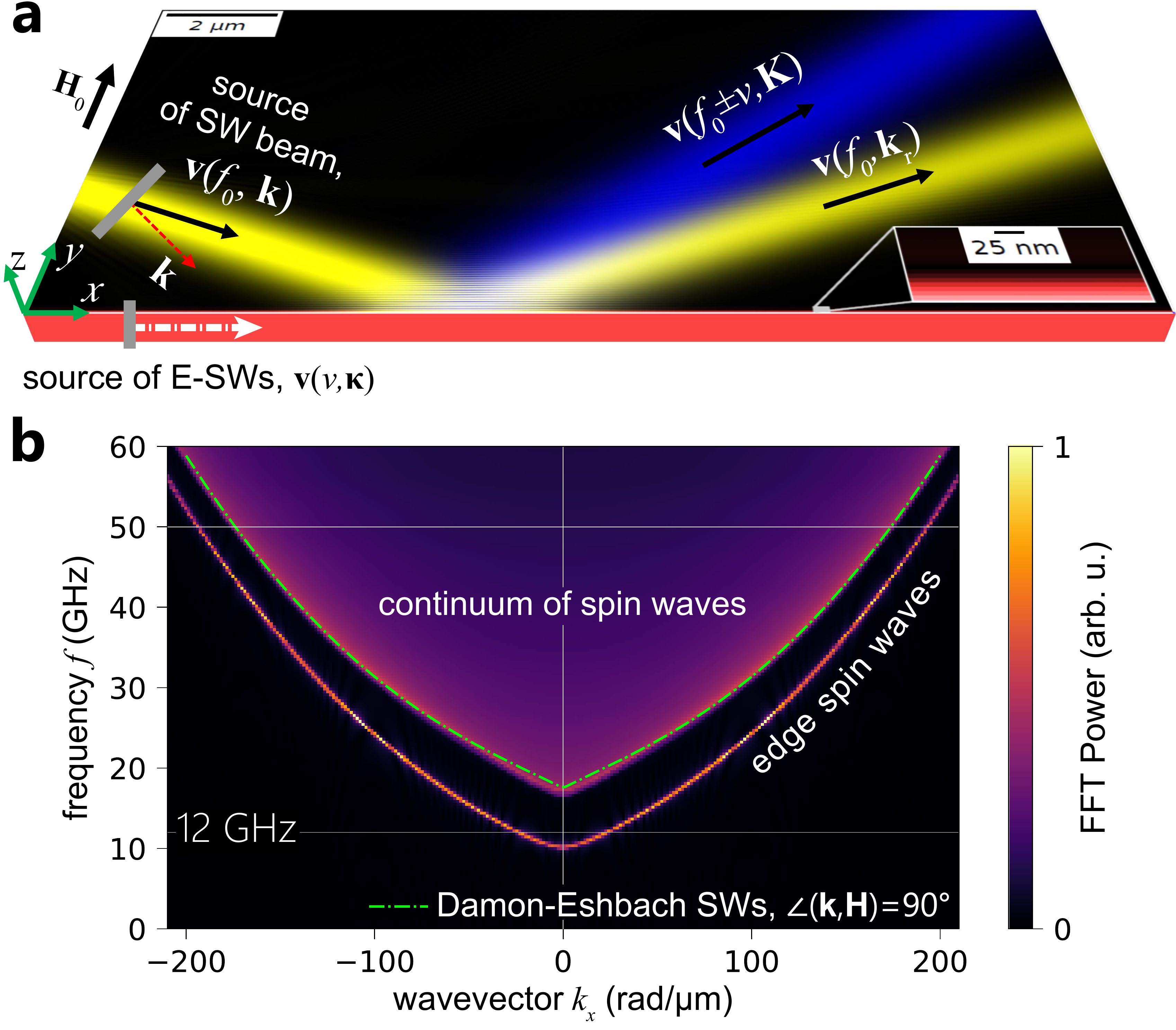}
\caption{\label{fig1_schema}
(a) Schema of the system under investigation. The red (at the edge), yellow, and blue (beams) colors correspond to the SWs at frequencies  $\nu$ (E-SW), $f$ (incident and reflected B-SW), and $f\pm\nu$ (inelastically scattered SWs), respectively. The white color denotes areas where at least two frequencies are mixed. The black arrows represent the group velocities related to the incident and scattered beams, whereas the dashed red arrow represents the wavevector of the incident wave. The white arrow represents the E-SW.
(b) Simulated dispersion relation (colormap in the background, where the color intensity corresponds to the averaged in space power related to the $x$ component of the wavevector) with well visible E-SW band at frequencies downshifted with respect to the continuum of the B-SW band (highlighted area). The dispersion of the B-SW mode with the wavevector perpendicular to the static magnetization, i.e., Damon-Eshbach mode, is marked by the green dash-dotted line.
}
\end{figure}

\section{Problem formulation}\label{Sec:problemFormulation}
%\section{Theoretical description}\label{Sec:model}

We consider a reflection of the SW beam from the permalloy Py (Ni$_{80}$Fe$_{20}$) film edge under  oblique incidence. The thin Py  film (the saturation magnetization $M_\mathrm{s}=800$~kA/m and the exchange constant $A=13$~pJ/m) of thickness $L=10$~nm is magnetized by the external magnetic field $\mathbf{H}_0=H_0\hat{\mathbf{y}}$ of the value $B_0 = \mu_0 H_0 = 0.3$~T (where  $\mu_{0}$ is the permeability of vacuum), directed perpendicularly to the film's edge, to the $(x,z)$ plane, see Fig.~\ref{fig1_schema}(a).   In this configuration, the sample is uniformly magnetized along the $y$-axis and the static demagnetizing field gradually decreases the value of the static component of the effective magnetic field $\mathbf{H}_{\mathrm{eff},y}$ at the vicinity of the film's edge, forming a suitable condition for a localization of the edge SWs\cite{McMichael2008,gruszecki2021_2harmonics}.

\subsection{Spin wave dynamics in thin films}

 In the micromagnetic approach the magnetization dynamics in a ferromagnetic film is described by the Landau-Lifshitz-Gilbert (LLG) equation of motion for the magnetization vector $\mathbf{M}$\cite{gilbert2004phenomenological}: 
\begin{equation}
\frac{\mathrm{d}\mathbf{M}}{\mathrm{d}t}=-\frac{\left|\gamma\right|\mu_{0}}{1+\alpha^{2}}\mathbf{M}\times\mathbf{H}_{\mathrm{eff}}-\frac{\alpha\left|\gamma\right|\mu_{0}}{M_{\mathrm{S}}\left(1+\alpha^{2}\right)}\mathbf{M}\times\left(\mathbf{M}\times\mathbf{H}_{\mathrm{eff}}\right),\label{eq:LLE}
\end{equation}
where $\alpha$ is the damping parameter, $\gamma$ is the gyromagnetic ratio and $\mathbf{H}_{\mathrm{eff}}$ is the effective magnetic field. The first term on the right side of the LLG equation describes the precession  of the magnetization around the effective magnetic field direction, whereas the second term introduces damping to that precession. In our study, the effective magnetic field consists of the external magnetic field $\mathbf{H}_{\mathrm{0}}$, the exchange field $\mathbf{H}_{\mathrm{ex}}$ and the magnetostatic field $\mathbf{H}_{\mathrm{m}}$: $\mathbf{H}_{\mathrm{eff}}=\mathbf{H}_{\mathrm{0}} + \mathbf{H}_{\mathrm{ex}}+\mathbf{H}_{\mathrm{m}}$.

We use the open-source software MuMax3\cite{vansteenkiste2014design} to solve numerically Eq.~(\ref{eq:LLE}) and simulate an inelastic scattering of the SW beam from the Py film edge.  The details of simulations including beam generation and post-processing are presented in  Appendix~\ref{Sec:AppSimulations}.

\subsection{Dispersion relation in thin ferromagnetic film}

In the case of a uniformly in-plane magnetized thin film of a thickness $L$, under linear approximation and free boundary conditions on the dynamic components of the magnetization at its surfaces, the dispersion relation for the SW propagating in the film plane is given by\cite{kalinikos1986theory,stancil2009spin}:
\begin{equation}
f(\mathbf{k})=\frac{1}{2\pi}\sqrt{\omega_{\mathrm{0}}\left(\omega_{\mathrm{0}}+\omega_{\mathrm{M}}F\left(\varphi,k\right)\right)},\label{eq:Dispersion}
\end{equation}
where $\mathbf{k}=(k_{x},k_{y})$ is the two-dimensional wavevector, $\omega_{\mathrm{M}}=\gamma\mu_{0}M_{\mathrm{s}}$, $\omega_{\mathrm{0}}=\left|\gamma\right|(B_0+ l_{\mathrm{ex}}^{2}\omega_{\mathrm{M}}k^{2})$, 
%\igorinline{in expression for $\omega_{\mathrm{0}}$ didn't explained what is $B_{\mathrm{0}}$}, 
$l_{\mathrm{ex}}=\sqrt{2A/(\mu_{0}M_{\mathrm{s}}^{2})}$  being the exchange length,  and $\varphi$ is the angle  between the wavevector and the external magnetic field. The function $F(\varphi, k)$ is defined as:
\begin{eqnarray*}
    \begin{split}
    F(\varphi,k)  =  & 1-P\left(kL\right)\cos^{2}\varphi  \\
     & +  \frac{M_{\text{s}}P\left(kL\right)\left[1-P\left(kL\right)\right]}{B_{0}\mu_0^{-1}+M_{\text{s}} l_{\mathrm{ex}}^{2}k^{2}}\sin^{2}\varphi,
  \end{split}
\end{eqnarray*}
where  $P\left(\xi\right) = 1-\left[1-\mathrm{exp}(-\xi)\right]/\xi$.

Using Eq.~(\ref{eq:Dispersion}), one may find isofrequency contours of the dispersion relation that are slices of the dispersion relation at a particular SW frequency over the $(k_{x},k_{y})$ wavevector plane. These isofrequency contours will be used in the analysis of the allowed wavevectors of the scattered waves.\cite{lokk2008isofrequency}

The direction of the wave energy flow is defined by the direction of the  group velocity $\mathbf{v}_{\mathrm{g}}$:
\begin{equation}
\mathbf{v}_{\mathrm{g}}(\mathbf{k})=\nabla_{\mathbf{k}}\omega(\mathbf{k}),
\end{equation}
where $\omega(\mathbf{k})=2 \pi f(\mathbf{k})$ is the dispersion relation defined in Eq.~(\ref{eq:Dispersion}) and $\nabla_{\mathbf{k}}=\left[\partial/\partial k_{x},\partial/\partial k_{y}\right]$ is a gradient in the wavevector space. Therefore, the direction of the SW group velocity is  normal to the isofrequency contour for a given $\mathbf{k}$. Importantly, the direction of the group velocity can be equated with the direction of propagation of the SW beam, i.e., its ray. For an isotropic media  the isofrequency contours are circular and  $\mathbf{v}_{\mathrm{g}}$ is parallel to $\mathbf{k}$, whereas, for anisotropic media, a typical situation for SWs in an in-plane magnetized thin film, the SW beam may propagate in a different direction than its wavevector points.

For the film thickness $L=10$ nm and the considered frequency range the variation of the dynamical components of the magnetization along the $z$-axis is negligible. 

\subsection{Spin waves in semi-infinite thin film}
The simulated dispersion relation, i.e., the SW frequency vs. the $x$ component of the wavevector, is presented in Fig.~\ref{fig1_schema}(b). The dispersion relation shows the continuum of B-SW band starting at frequency $f \sim$ 20~GHz (highlighted area) and an isolated SW band at smaller frequencies starting at $ \sim$ 10.8~GHz. The edge of the continuum B-SW for large $k_x$ values is determined by the Damon-Eshbach mode, it is when $\mathbf{k}\perp  \mathbf{H}_0$ (propagating far from the edge), and it is marked by the green dash-dotted line. At a small $k_x$ the edge of the band is determined by the waves propagating along the magnetic field direction.\cite{stancil2009spin}

The isolated band at downshifted frequencies in  Fig.~\ref{fig1_schema}(b) corresponds to an SW localized at the film's edge that propagates along the edge with the wavevector $\mathbf{\kappa}\perp  \mathbf{H}_0$. Its confinement, together with the lowered frequencies with respect of B-SW, enables a selective, and independent excitation of the B-SW beam at $f_0 >20$~GHz and E-SW at $\nu \in (10.8, 20)$~GHz, i.e., at different frequencies in the same system. Thus, it is a well suitable system to answer the basic question of this study: Is it possible to scatter the incident B-SW beam on the E-SWs and excite B-SW beams at frequencies $f_0 \pm \nu$, as it is schematically shown in  Fig.~\ref{fig1_schema}(a)? We answer this question, firstly developing an analytical model of the inelastic B-SW scattering on E-SW, and subsequently conducting detailed micromagnetic simulations.

\section{Model of inelastic scattering of spin waves \label{Sec:model}}
We have shown that in our system the two kinds of SW excitations can exist (Fig.~\ref{fig1_schema}): 1) the plane B-SWs, incident and scattered, in the bulk of the film, and (2) the E-SW localized at the film edge. This allows us to consider the nonlinear scattering of B-SWs on E-SWs as a three-magnon process and, therefore, to develop the analytical model.

%We denote the amplitudes of the localized and plane SWs as $a_k$, where the index $k=\kappa$ for the E-SW, and with $k = \mathbf{k}$ for a B-SW. 

We describe the unit magnetization vector $\mathbf{m}(\Theta,\Phi) = \mathbf{M}/M_\mathrm{s}$ by the spherical angles $\Theta$ and $\Phi$. Then, the magnetic free energy is $W=\int_V dV w$,  with the energy density $w$ integrated over the ferromagnet volume $V$, is defined in the following form:
\begin{equation}
w= A \Big[ (\nabla \Theta )^2 + \mathrm{sin}^2\Theta (\nabla \Phi )^2 \Big]  - \mu_{0}\mathbf{M}\cdot\mathbf{H}_0-\frac{\mu_{0}}{2}\mathbf{M}\cdot\mathbf{H_\mathrm{m}}.
\label{eq:magneticEnergy}
\end{equation}
To study magnetization dynamics we use the Landau-Lifshitz equation in the form suggested by Slonczewski\cite{slonczewski1979}:
\begin{equation}
M_\mathrm{s}\dot{\Phi} \mathrm{sin}\Theta =\gamma\left(\frac{\delta w}{\delta \Theta}\right),~~~~
M_\mathrm{s}\dot{\Theta} \mathrm{sin}\Theta =-\gamma\left(\frac{\delta w}{\delta \Phi}\right)
\label{eq:LLeq_Slonczewski},
\end{equation}
where an overdot denotes the derivative with respect to time.

The SWs are defined as small perturbations on the magnetization background (in general, the magnetization background can be inhomogeneous and dynamical). The magnetization can be written as a sum of the static $\mathbf{m}_0$ and a small   SW contribution $\mathbf{m}_\mathrm{\text{sw}}$: $\mathbf{m}=\mathbf{m}_0 + \mathbf{m}_\mathrm{\text{sw}}$. Analogously, the SW angles are defined as small deviations from the equilibrium magnetization angles $\Theta=\Theta_0+\vartheta$, and $\Phi=\Phi_0 + \psi$. We assume that the equilibrium magnetization is parallel to the bias magnetic field and perpendicular to the film edge plane, thus $\mathbf{m}_0(\Theta_0, \Phi_0)= \hat{\mathbf{y}}$,  $\Theta_0=\pi/2$ , and $\Phi_0 = -\pi/2$ (see, Fig.~\ref{fig1_schema}a). The components of the SW magnetization in quadratic approximation are $\mathbf{m}_\mathrm{sw}=(\psi,\vartheta ^2/2+\psi^2/2,-\vartheta)$.
%,  are substituted to Eqs.~(\ref{eq:LLeq_Slonczewski}). 

The inelastic scattering of incident/reflected plane B-SWs by the E-SW is determined by an interaction Hamiltonian $H_\mathrm{int}$. It contains cubic 
%($aaa$) 
terms of the SW amplitudes and other high-order interaction terms. 
%There is a lot of the papers considering the cubic and quartic terms in the SW energy decomposition in relation to the problem of non-linear ferromagnetic resonance, parallel SW pumping, and SW relaxation, see review in the textbooks by Gurevich and Melkov\cite{gurevich1996} and Stancil et al. \cite{stancil2009spin}. 
We follow below the calculation scheme first suggested by Suhl\cite{suhl1957}, who considered the problem of the parallel SW pumping in a bulk sample, as an interaction of the uniform precession magnon with the finite-wavevector SWs.

Accounting the decompositions $\mathbf{m}=\mathbf{m}_0+ \mathbf{m}_\mathrm{sw}$ and $\mathbf{H}_\mathrm{m}= \mathbf{H}_{\mathrm{m}}^0+ \mathbf{h}_\mathrm{m}$ into the static and dynamical parts, the system of Eqs.~(\ref{eq:LLeq_Slonczewski}), keeping only the linear and quadratic terms in the SW magnetization angles, becomes
\begin{equation}
\begin{split}
\frac{1}{\gamma}\dot{\psi}(\boldsymbol{\varrho}) = & -\frac{2A}{M_s}\nabla^2\vartheta(\boldsymbol{\varrho})-\mu_{0}H_{\mathrm{m},y}^0\vartheta(\boldsymbol{\varrho})-\\ 
&-\mu_{0}M_\mathrm{s} \int d\boldsymbol{\varrho}^\prime G_{zz}\vartheta (\boldsymbol{\varrho}^\prime) \\
		& -\mu_{0}M_\mathrm{s} \vartheta(\boldsymbol{\varrho}) \int d\boldsymbol{\varrho}^\prime G_{yx} \psi (\boldsymbol{\varrho}^\prime), \\
-\frac{1}{\gamma}\dot{\vartheta}(\boldsymbol{\varrho}) = &-\frac{2A}{M_s}\nabla^2\psi(\boldsymbol{\varrho})-\mu_{0}H_{\mathrm{m},y}^0\psi(\boldsymbol{\varrho}) \\
&-\mu_{0}M_\mathrm{s} \int d\boldsymbol{\varrho}^\prime G_{xx}\psi (\boldsymbol{\varrho}^\prime)\\
		& -\frac{\mu_{0}}{2} M_\mathrm{s} \int d\boldsymbol{\varrho}^\prime G_{xy} (\vartheta^2(\boldsymbol{\varrho}^\prime)+\psi^2(\boldsymbol{\varrho}^\prime)) \\
		&- \mu_{0}M_\mathrm{s} \psi(\boldsymbol{\varrho}) \int d\boldsymbol{\varrho}^\prime G_{yx}\psi (\boldsymbol{\varrho}^\prime),
\end{split}
\label{eq:LLeq_Slonczewski_v2}
\end{equation}
where $G_{\alpha\beta} \equiv G_{\alpha\beta}(\boldsymbol{\varrho}, \boldsymbol{\varrho}^\prime)$ are the magnetostatic Green functions averaged over the film thickness and defined by Eqs.~(19)-(21) of Ref.~[\onlinecite{guslienko2011}], and $\boldsymbol{\varrho}=(x,y)$, $\boldsymbol{\varrho}^\prime=(x\prime, y\prime)$ are the position vectors. The symbol $H_{\mathrm{m},y}^0=H_{\mathrm{m},y}^0\left(y\right)$ represents the static dipolar field within the film, averaged over the $x$-, and the $z$-coordinates. The field strongly depends on the coordinate $y$ only near the film edge, in the region with the width about of the film thickness. 

It is evident from Eqs.~(\ref{eq:LLeq_Slonczewski_v2}) that neither the exchange nor Zeeman interaction do not contribute to the quadratic terms in the equations of motion. The quadratic terms are responsible for the interaction between the SWs and they arise due to the non-local magnetostatic coupling. The interaction energy density $w_\mathrm{int}=-M_\mathrm{s} \mathbf{h}_\mathrm{m}\cdot\mathbf{m}_\mathrm{sw}$ can be explicitly written as 
\begin{equation}
\begin{split}
w_\mathrm{int}=-\frac{\mu_{0}}{2}M_\mathrm{s}^2 \Bigg[& \psi \int d\boldsymbol{\varrho}^{\prime} G_{xy}\left(\vartheta^2(\boldsymbol{\varrho}^{\prime})+\psi^2(\boldsymbol{\varrho}^{\prime})\right)\\
&+(\vartheta^2+\psi^2) \int d\boldsymbol{\varrho}^{\prime} G_{yx}\psi(\boldsymbol{\varrho}^{\prime}) \Bigg].
\end{split}
\end{equation}

Finally, using the symmetry properties of the Green functions $G_{\alpha\beta}(\boldsymbol{\varrho},\boldsymbol{\varrho}^\prime) =G_{\beta\alpha}(\boldsymbol{\varrho},\boldsymbol{\varrho}^\prime)$, the cubic interaction energy can be written in the form
\begin{equation}
\begin{split}
& W_\mathrm{int}=\\
&-\mu_{0}M_\mathrm{s}^2L  \int d\boldsymbol{\varrho} \Big[ \psi(\boldsymbol{\varrho}) \int d\boldsymbol{\varrho}^{\prime} G_{xy} \left( \vartheta^2(\boldsymbol{\varrho}^{\prime})+\psi^2(\boldsymbol{\varrho}^{\prime})\right) \Big].
\end{split}
\label{eq:energy_5}
\end{equation}

In general, we can decompose the complex dynamic magnetization $\Psi=\vartheta+i\psi$ in a series of the eigenfuctions $m_k\left(\boldsymbol{\varrho}\right)$  of the linear eigenvalue problem given by the linear part of the system of Eqs.~(\ref{eq:LLeq_Slonczewski_v2}) 
\begin{equation}
\Psi\left(\boldsymbol{\varrho},t\right)=\sum_{k} a_k\left(t\right)m_k\left(\boldsymbol{\varrho}\right),
\end{equation}
where $a_k$ is the SW eigenmode amplitude, then the SW interaction energy~(\ref{eq:energy_5}) can be written in the form
\begin{equation}
\begin{split}
& W_\mathrm{int}=\\
&-\frac{1}{2i}\mu_{0}M_{\mathrm{s}}^{2} L 
\sum_{k,k^{\prime},k^{\prime \prime}} \Big[ B_{kk^{\prime}k^{\prime\prime}} a_{k}a_{k^{\prime}}a_{k^{\prime\prime}}^{*}-B_{kk^{\prime}k^{\prime\prime}}^{*}a_{k}^{*}a_{k^{\prime}}^{*}a_{k^{\prime\prime}} \Big],
\end{split}
\label{eq:energy_7}
\end{equation}
with the SW scattering amplitudes 
\begin{equation}
B_{kk^{\prime} k^{\prime\prime}}=
\int  d^{2} \boldsymbol{\varrho} \int d^{2} \boldsymbol{\varrho}^{\prime} G_{xy} 
 m_{k}(\boldsymbol{\varrho})
  m_{k^{\prime}} (\boldsymbol{\varrho}^{\prime}) m_{k^{\prime \prime}}^{*}(\boldsymbol{\varrho}^{\prime}). 
  \label{eq:energy_8}
\end{equation}
The first term in Eq.~(\ref{eq:energy_7}) corresponds to the elementary SW confluence process, whereas the second one corresponds to the SW splitting process. Let us investigate in more detail the confluence process described by the amplitude given by Eq.~(\ref{eq:energy_8}). The SW splitting amplitude is just complex conjugate to the confluence amplitude. 

In the system under investigation, there is a translation invariance along the film edge. Therefore, the wavevector component along the $x$-axis can be introduced. According to markings in Fig.~\ref{fig1_schema}, the index $k^\prime=\kappa$ will describe the localized E-SW, $m_{\kappa}\left(x,y\right)=m_{\kappa}\left(y\right) \mathrm{exp}\left(i\kappa x\right)$, the non-localized B-SW $m_\mathbf{k}\left(x,y\right)=m_\mathbf{k}\left(y\right)\mathrm{exp}\left(ik_xx\right)$ are described by the indices $\mathbf{k}$, and $\mathbf{K}$, for the incident and inelastic scattered wave, respectively. Although, the linear system of Eqs.~(\ref{eq:LLeq_Slonczewski_v2}) cannot be solved analytically, we can assume that the non-localized B-SWs described by the indices $\mathbf{k}$ and $\mathbf{K}$ can be approximated by the plane waves $m_k\left(\boldsymbol{\varrho}\right)=\mathrm{exp}\left(i\mathbf{k}\cdot \boldsymbol{\varrho}\right)$ with the in-plane wavevectors $\mathbf{k}$ and $\mathbf{K}$. Let us use explicitly the SW eigenfunction dependence on $x$. The amplitude of the scattering from $\mathbf{k}$-state to $\mathbf{K}$-state via a localized $\boldsymbol{\kappa}$-state, i.e., absorbing the E-SW, is
\begin{equation}\
\begin{split}
&B_{\mathbf{k}\boldsymbol{\kappa}\mathbf{K}}=
2 \pi \int_{0}^{\infty}  dy dy^{\prime}\\
&\times\Big[ g_{xy}(y,y^{\prime};-k_{x})  
m_{\mathbf{k}}(y)m_{\boldsymbol{\kappa}}(y^{\prime}) m_{\mathbf{K}}^{*}(y^{\prime})
\delta (k_{x}+\kappa-K_{x}) \Big],\label{Eq:Bkkk}
\end{split}
\end{equation}
where $\delta(\xi)$ is the Dirac delta function and the simplified Green function $g_{xy}$ is
\begin{equation}
\begin{split}
&g_{xy}(y,y^{\prime};k)=\\
&-\frac{1}{2 \pi} \int_{-\infty}^{\infty} dp \frac{kp}{k^2+p^2}P\left( L\sqrt{k^2+p^2}\right) e^{ip(y-y^{\prime})}.
\end{split}
\label{eq:simplGreenFunc_9}
\end{equation}

The amplitude given by Eq. (\ref{Eq:Bkkk})
% , where $f(x) = 1-\left[1-\mathrm{exp}(-x)\right]/x$, \pawelinline{$f(x)$ is already defined as P(k)}
is calculated without any assumption about the $y$-dependence of the SW eigenfunctions. If we assume additionally that $m_\mathbf{k}\left(y\right)=\mathrm{exp}\left(ik_yy\right)$, then the scattering amplitude Eq.~(\ref{Eq:Bkkk}) is transformed to
\begin{equation}
\begin{split}
B_{\mathbf{k}\boldsymbol{\kappa}\mathbf{K}}=&-2 \pi \delta(k_{x}+\kappa-K_{x}) \frac{k_x k_y}{ k^2}P(kL)  \\
&\times \int_{0}^{\infty} dy m_{\boldsymbol{\kappa}}(y)e^{i(k_{y}-K_{y})y}.
\end{split}
\label{eq:scatteringAmplitude_10}
\end{equation}

The E-SW mode profiles $m_{\kappa}\left(y\right)$, in principle, can be obtained from a solution of the linearized system of Eqs.~(\ref{eq:LLeq_Slonczewski_v2}) accounting for the static and dynamic dipolar fields. However, this is practically impossible to implement analytically (see, Refs.~[\onlinecite{gubbiotti2004,bayer2005, lara2017information}]). Therefore, we use a trial function for the low frequency edge eigenmode, $m_{\kappa}\left(y\right)$, in the form $m_\kappa\left(y\right)=\mathrm{exp}\left(-y/\Delta \right)$ and assume that the localization width $\Delta$ of the E-SW is about of several tens of nm. The drawback of this trial function is that it is not orthogonal to the plane SWs. Then, Eq.~(\ref{eq:scatteringAmplitude_10}) yields to the scattering amplitude
\begin{equation}
\begin{split}
&B_{\mathbf{k}\boldsymbol{\kappa}\mathbf{K}}=\\
&-2 \pi \Delta \delta(k_{x}+\kappa-K_{x}) \frac{k_x k_y}{ k^2}P(kL)  \times \chi \left( (k_{y}-K_{y})\Delta\right),
\end{split}
\label{eq:scatteringAmplitude_11}
\end{equation}
% \pawelinline{\textbf{Q}: Konstantin, do the units agree? }
where the function $\chi\left(x\right)=\left(1+ix\right)/\left(1+x^2\right)$ reflects non-conservation of the $y$-component of the SW momentum due to breaking the translation invariance along the $y$-axis by the film edge.

If the SW beam is elastically reflected from the film edge and the Snell law is satisfied, we can accept that $k_{y}=-k_{\textrm{r},y}$ and $k_{y}-k_{\textrm{r},y}=2k_{y}$, but for the inelastic scattering the $y$-component of the scattered wave, $K_{y}$, needs to be extracted from the dispersion relation, Eq.~(\ref{eq:Dispersion}). 
The SW confluence (absorption of E-SW by B-SW) amplitude given by Eqs.~(\ref{eq:scatteringAmplitude_10}) and~(\ref{eq:scatteringAmplitude_11}) is represented mainly by the incident SW wavevector components. The SW splitting (emission of E-SW by B-SW) amplitude is given by Eqs.~(\ref{eq:scatteringAmplitude_10}),~(\ref{eq:scatteringAmplitude_11}), where one has to substitute $\kappa$ to $- \kappa$ and the function $\chi(x)$ to its complex conjugate. Note that the amplitude angular dependence is 
% $B_{\mathbf{k}\boldsymbol{\kappa}\mathbf{K}}\propto \mathrm{sin}2\varphi_{\mathbf{k}}F(2k\Delta\mathrm{sin}\varphi_{\mathbf{k}})$, where $\varphi_{\mathbf{k}}$ is counted form the $x$-axis, i.e., the film edge plane $(x,z)$.  
% $B_{\mathbf{k}\boldsymbol{\kappa}\mathbf{K}}\propto \mathrm{sin}(2\varphi) \chi(2k\Delta\mathrm{cos}\varphi)$, where $\varphi$ is counted form the $y$-axis. 
$B_{\mathbf{k}\boldsymbol{\kappa}\mathbf{K}}\propto \mathrm{sin}(2\varphi) \chi( (k_{y}-K_{y})\Delta )$, where $\varphi$ is already defined angle of incidence being counted from the $y$-axis, $k_y<0$ and $K_y>0$. 
The function $B_{\mathbf{k}\boldsymbol{\kappa}\mathbf{K}}(\varphi)$ reaches its maximum at $\varphi=\pi/4$ if the E-SW mode localization is strong with respect to the wavelengths of incident and scattered waves, as described by the following condition

\begin{equation}
(k_y-K_y) \Delta \ll 1. \label{Eq:localization_condition}
\end{equation}

The scattering amplitude becomes negligibly small in the opposite limiting case of weakly localized E-SW, $(k_y-K_y) \Delta \gg 1$. 
% \pawelinline{Konstantin, we have introduced here some changes, check if you agree}
%\konstantininline{Note the change of the angular dependence.
%\pawelinline{I have replaced $\mathrm{sin}{\varphi_\mathbf{k}}$ by  $\mathrm{cos}{\varphi}$ with already defined $\varphi$ in Eq.~\ref{eq:Dispersion}} }.

The scattering amplitude $B_{\mathbf{k}\boldsymbol{\kappa}\mathbf{K}}(\varphi)$ accounts explicitly for the momentum conservation law
\begin{equation}
K_{x,\pm}=k_{x} \pm \kappa \label{Eq:Conserw_momentum}
\end{equation}
along the film edge. From the other side, the energy conservation law 
\begin{equation}
\omega_\mathbf{{K},\pm}=\omega_{\mathbf{k}} \pm \omega_{\boldsymbol{\kappa}} \label{Eq:Conserw}
\end{equation}
should be satisfied for the confluence ($+$) and splitting ($-$) processes, where $\omega_{\mathbf{k}}=2\pi f_0$ and $\omega_{\boldsymbol{\kappa}}=2\pi \nu$. Therefore, the confluence and splitting  amplitudes depend on the final plane B-SW wavevector $\mathbf{K}$  and the $x$-component of the E-SW vector, $\boldsymbol{\kappa}$. Assuming the plane wave approximation $m_k\left(\boldsymbol{\varrho}\right)=\mathrm{exp}\left(i\mathbf{k}\cdot\boldsymbol{\varrho}\right)$ for the non-localized B-SW and neglecting the static dipolar field $H_{\mathrm{m},y}^{0}$, the linearized system of Eqs.~(\ref{eq:LLeq_Slonczewski_v2}) can be essentially simplified yielding the well-known SW dispersion relation of an infinite film with thickness $L$, defined in Eq.~(\ref{eq:Dispersion}).

%\pawelinline{It seems to me that it is worthwhile here to draw some conclusions from the prepared analytical model. This would be particularly helpful in the context of its confrontation with simulations.}
% \cite{kalinikos1986theory}

% \begin{equation}
% \begin{split}
% \frac{1}{\gamma^2}\omega_\mathbf{k}^2=&\left[\frac{2A}{M_s}k^2+H+4\pi M_{\mathrm{S}}\left(1-f\right)\right]\\
% \times& \left[\frac{2A}{M_{\mathrm{S}}}k^2+H+4\pi M_{\mathrm{S}} f \mathrm{sin}^2\varphi_\mathbf{k}\right],
% \end{split}
% \end{equation}

% where $f=f\left(kL\right)$, and $H=-H_y>0$ is in-plane bias magnetic field. 
% \pawelinline{Do we need here dispersion relation that is already defined?}

\section{Results and discussion}\label{Sec:results}
\subsubsection{Analysis of B-SW beam scattering by E-SW}

\begin{figure*}[!t]
\includegraphics[width=15cm]{ 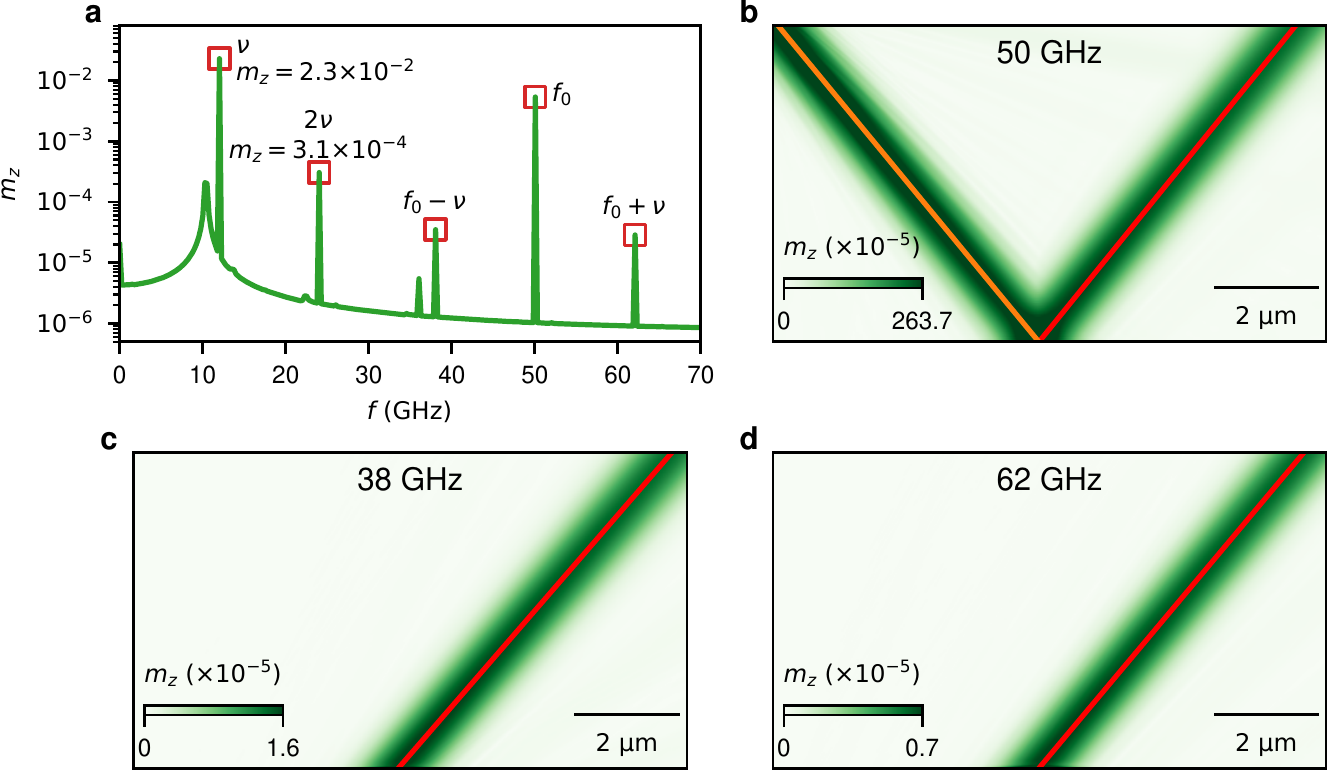}
\caption{(a) The spectrum of SWs with the most prominent frequency peaks corresponding to the E-SWs ($\nu$), 2nd harmonic generated SWs ($2\nu=24$~GHz), SW beam of frequency $f_0=50$~GHz, and the inelastically scattered SW beams $f_0 \pm \nu$ at the frequencies of 38 and 62~GHz.   (b) The intensity map of the incident and reflected SW beams at frequency 50~GHz. 
%\konstantininline{where are the antennas in Fig.2b?} 
(c) and (d) Intensity maps of SWs at frequency 38~GHz and 62~GHz, respectively. The angles of propagation of the scattered beams in (c) and (d) differ slightly with respect to each other and with respect to the primary reflected beam presented in (b).
The locations of antennas exciting both, SW beam and E-SWs, are schematically shown in Fig.~\ref{fig1_schema}(a).
} \label{fig2_beams}
\end{figure*}

\begin{figure*}[!t]
\includegraphics[width=14cm]{ 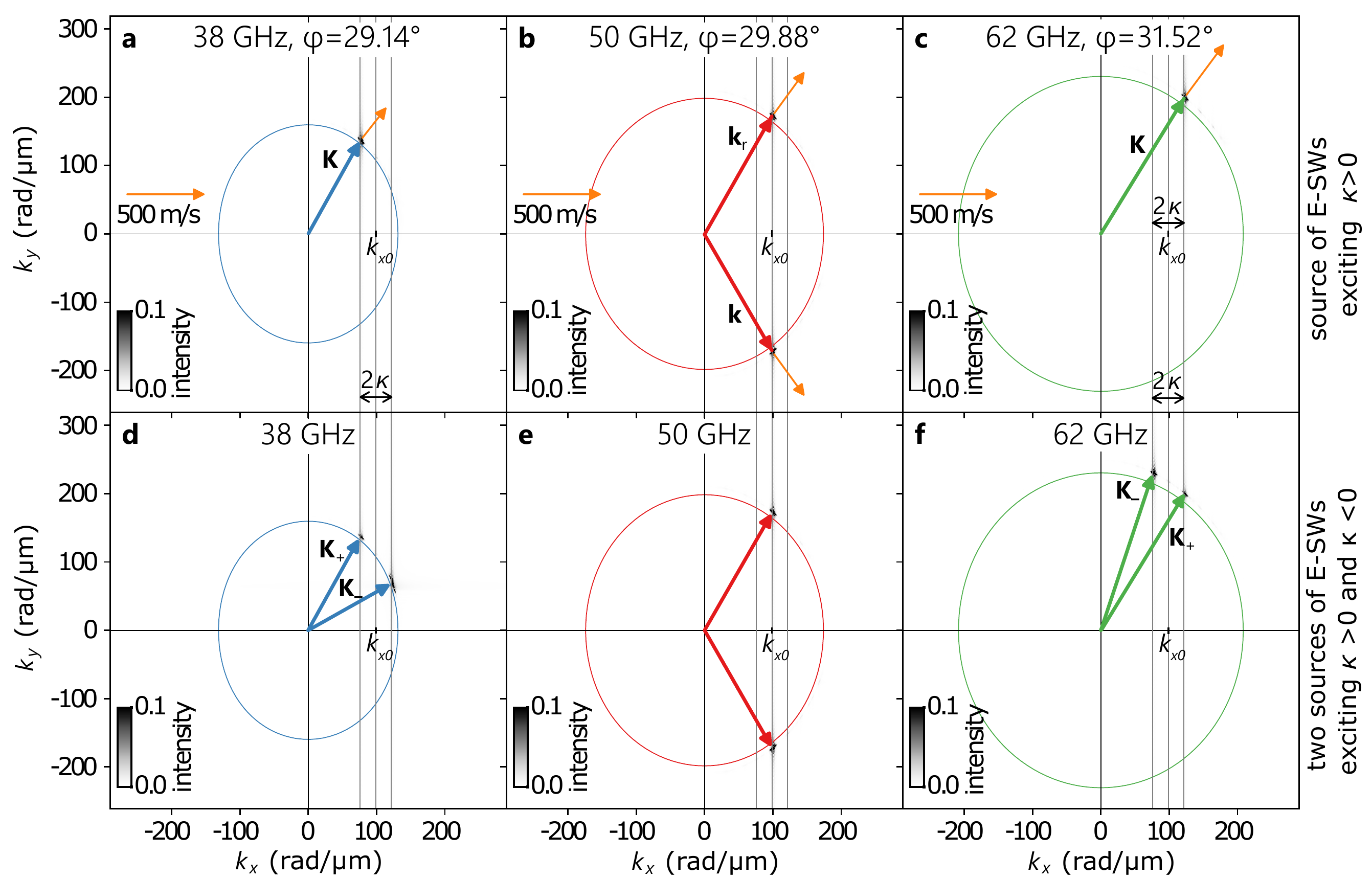}
\caption{
The momentum-space representation of the ISSW. The grayscale color map in the background represents the profile of SWs in the k-space obtained from simulations. The sharp black regions that are marked by the color arrows represent the wavevectors of the SW beams.
In (a)-(c) are shown results obtained for the case from Fig.~\ref{fig2_beams}, where the source of E-SW is placed on the left side of the incident beam spot, whereas in (d)-(f) there are two sources of E-SWs located on both sides of the incident beam spot. In (a) and (c) are shown results for $f_0-\nu = 38$~GHz, in (b) and (e) for 50~GHz -- specular reflection, whereas in (c) and (f) for $f_0+\nu=62$~GHz. 
The solid ellipses represent isofrequency contours for corresponding frequencies resulting from the analytical dispersion relation, Eq.~(\ref{eq:Dispersion}). The thin-orange arrows correspond to the group velocities obtained analytically for a given wavevector (their length notes the magnitude of $\mathbf{v}_\mathrm{g}$). The angle $\varphi=\mathrm{tan^{-1}}(k_x/k_y)$ in (a)-(c) denotes the angle of the phase velocity with respect to the normal to the film edge.
}\label{fig:fig3_IFC}
\end{figure*}

%Let us consider a reflection of the SW beam obliquely propagating in a thin permalloy film (the saturation magnetization $M_\mathrm{S}=800$ kA/m, the exchange constant $A=13$ pJ/m) magnetized by the external magnetic field $\mathbf{H}$ of value $\mu_0 H=0.3$ T directed perpendicularly to the film's edge ($xOz$ plane), see Fig.~\ref{fig1_schema}(a).   In this scenerio, in the vicinity of film's edge, the static demagnetizing field gradualy decreases the value of the static effective magnetic field $\mathbf{H}_{\mathrm{eff},0}$\cite{gruszecki2014goos}.

To demonstrate the inelastic scattering of B-SWs on E-SW we conducted micromagnetic simulations. In Fig.~\ref{fig2_beams}(a)  we show the frequency spectrum calculated for the SW beam excited at the frequency $f_0=50$~GHz with the wavevector $\mathbf{k}$ directed under the angle of incidence $\varphi = 30^\circ$ that falls at the film's edge, where an independently excited the E-SW at frequency $\nu =12$~GHz  propagates along the $x$-axis with $\kappa>0$.

% To verify the possibility of ISSW on the E-SWs, we have perfomed micromagnetic simulations to scatter the incident SW beam at the E-SWs and excite secondary spin wave beams with shifted frequencies, let us analyze results of micromagnetic simulations performed by mumax3\cite{vansteenkiste2014design} (All the technical details of micromagnetic simulations can be found in  Appendix~\ref{Sec:AppSimulations}). In the simulations, we have assumed that the incident SW beam of frequency $f_0=50$~GHz and wavevector $\mathbf{k}$ directed under the angle of incidence 30$^\circ$  falls at the film's edge where E-SWs of frequency $\nu =12$ ~GHz are localized in. 

The first signature of ISSW is the presence of a series of peaks in Fig.~\ref{fig2_beams}(a). The two most pronounced peaks correspond to the frequencies $f_0=50$~GHz and $\nu = 12$~GHz, i.e., to the excited and elastically reflected B-SW, and E-SW, respectively. A small, very broad peak at frequency 10.8 GHz corresponds to the resonance of the E-SW for $\kappa=0$, and it always appears after activating the antenna that excites E-SWs. There is also a peak at $2\nu=24$~GHz related to the second-harmonic generation by the E-SW. The process of second-harmonic generation of plane waves by the E-SW has been already described in detail in Refs.~[\onlinecite{gruszecki2021_2harmonics, hermsdoerfer2009}], here, however, we focus on an inelastic scattering. The two other peaks marked in Fig.~\ref{fig2_beams}(a), at  the frequencies $f_0 - \nu=38$~GHz and $f_0 + \nu=62$~GHz, correspond exactly to the ISSWs, accordingly with the energy conservation law, Eq.~(\ref{Eq:Conserw}). Namely, these peaks correspond to the splitting and confluence processes, respectively. Interestingly, the amplitude of the wave generated in the splitting process is greater than in the confluence process, pointing at a higher efficiency of the splitting than the confluence process.
Moreover, the occurrence of a peak at the $f_0-\nu$ frequency indicates, the emission of the E-SW at a frequency $\nu$ in the splitting process. It suggests that an already existing E-SW at frequency $\nu$  triggers this process. Instead of having a three-SW splitting with random wave excited, we observe the assisted splitting process, also known in the literature as a stimulated splitting process\cite{wang2018probing, zhang2018}.

Let us analyze the SW beams shown in Fig.~\ref{fig2_beams}(b)-(d) corresponding to the detected frequencies. At the frequency $f_0$ there are two beams that propagate towards and outwards the interface, namely, the incident beam and the reflected beam [Fig.~\ref{fig2_beams}(b)]. 
These beams at frequencies $f_0\pm\nu$, presented in Fig.~\ref{fig2_beams}(c) and (d),  propagate only outward the film's edge, thus, confirming their excitation at the film's edge as a result of the ISSW.
In Fig.~\ref{fig2_beams}(b)-(d), we plot also the beam rays (solid lines along the beams), which are associated with the direction of the group velocity.
All the beams propagate under very close angles, namely, the elastically reflected beam at frequency 50~GHz propagates under the angle of $39^\circ$ with respect to the normal to the film edge, whereas the scattered beams at frequencies 38~GHz and 62~GHz, propagate under the angles of $41^\circ$ and  $40^\circ$, respectively. The difference in the beam ray directions of incident and reflected waves arises from the anisotropy of the isofrequency contour -- see, the elliptical shape of the contours in Fig.~\ref{fig:fig3_IFC}. Hence, also the directions of the phase velocity (30$^\circ$) and group velocity (39$^\circ$) differ from each other.

In order to detect the values of wavevectors corresponding to the SWs and verify whether the momentum conservation law, Eq.~(\ref{Eq:Conserw_momentum}), is fulfilled, we  calculate spatial Fast Fourier Transform (FFT) for SWs at frequencies $\nu$, $f_0$, and $f_0 \pm \nu$. For the frequency $\nu=12$~GHz we obtain the wavevector of value $\boldsymbol{\kappa}=\mathbf{\hat{x}} 2\pi/(285 \text{ nm})$, i.e., the wavelength equal to 285 nm.  The $k$-space maps of the FFTs for the other frequencies are shown in Fig.~\ref{fig:fig3_IFC}(a)-(c) by a grayscale colormaps in the background, there are the black spots corresponding to the considered SWs.

The  $k$-space representation  at the frequency $f_0$ [Fig.~\ref{fig:fig3_IFC}(b)] shows the two peaks corresponding to the incident and reflected beams, propagating with the negative and positive $k_y$, respectively. These two waves have the same tangential to the interface component of the wavevector, $k_x$, in agreement with the momentum conservation law for the wave reflection from the edge possessing translational symmetry.
%Snell's law for SW reflection\cite{stigloher2016snell}. 
On the other hand, at $f_0\pm\nu$ [Fig.~\ref{fig:fig3_IFC}(a) and (c)] there are visible single peaks at the positive $k_y$ half-space. 
The tangential components of their wavevectors, $K_x$ are shifted by $\kappa$ with respect to $k_x$, in agreement with  the momentum conservation laws defined in Eq.~(\ref{Eq:Conserw_momentum}).
% namely $K_x=k_{x,f_0+\nu}+\kappa$ and $K_x=k_{x,f_0-\nu}-\kappa$ for confluence and split, process,, respectively.
Accordingly with the analytical predictions of Sec.~\ref{Sec:model}, we confirm that for $f=f_0 \pm \nu$ we get the scattered beam at the wavevector $\mathbf{K}_{\pm}=(k_{x} \pm \kappa, K_y)$, with the $K_y$-component allowing 
to satisfy the dispersion relation of the Py film $f(\mathbf{K})$,
% for satisfying by  $f(\mathbf{K})$ the SW dispersion relation of the Py film,
i.e., $\mathbf{K}$ pointing at the isofrequency contour, exactly as demonstrated in Fig.~\ref{fig:fig3_IFC}(a) and (c). Interestingly, for the cases shown in Fig.~\ref{fig:fig3_IFC}(a)-(c), the mode profiles in $k$-space reveal that the angles of the phase velocities for the reflected and the two scattered beams are almost identical, and are 29.9$^\circ$, 29.1$^\circ$, and 31.5$^\circ$, respectively.

The demonstration presented above opens up also other ways to use the three-magnon ISSW process. Namely, depending on the location of the source of E-SW in relation to the beam spot, i.e., the change of the $\kappa$ sign, we can obtain a scattered beam propagating at different angles. Moreover, if we use a source of E-SW at both sides of the beam spot exciting a standing E-SW, we can obtain the two pairs of scattered beams at frequencies $f_0 + \nu$ and $f_0 - \nu$. 
Confirmation of such a scenario is displayed in Figs.~\ref{fig:fig3_IFC}(d), (e), and (f) where we show the SWs in the $k$-space at frequencies 38~GHz, 50~GHz, and 62~GHz, respectively, obtained from the scattering of the B-SW beam at 50~GHz on 12~GHz standing E-SW. It is visible, that at 38~GHz [Fig.~\ref{fig:fig3_IFC}(d)] and 62~GHz [Fig.~\ref{fig:fig3_IFC}(f)] the pair of black spots are present. The spots from the pair correspond to the two scattered beams propagating under different angles, which are determined by the momentum conservation Eq.~(\ref{Eq:Conserw_momentum}) for the $x$-component $K_{x,\pm}=k_x\pm \kappa$, and  the $y$-components $K_{y}$  matching with the isofrequency contour, independently at $f_0+\nu$ and $f_0 - \nu$. Interestingly, the effect can be considered as a splitting of the ISSW into two beams.

\subsubsection{Utilizing ISSW -- SW demultiplexing}

The $\mathbf{K}_\pm =(k_{x} \pm \kappa, K_y)$ dependence combined with the anisotropy of the SW dispersion, which varies with the frequency, makes it possible to obtain the frequency-dependent direction of the inelastically scattered SW beams. 
Such a behaviour can be used to spatially separate information carried by E-SWs at different frequencies, i.e.,  useful for SW demultiplexing or to design an SW nanoscale spectrum analyzer.\cite{papp2017nanoscale}  To achieve that, the angle of the inellastically scattered beam should strongly depend on the frequency of E-SW.

%For example, as in the result shown in Fig.~\ref{fig2_beams}, the rays of the two scattered beams are almost identical with the ray of the reflected beam. However, the opposite effect can also be obtained, i.e., the beams propagating at significantly different angles. 
%Our goal is to obtain beams propagating at significantly different angles to ensure maximum good spatial separation of waves at different frequencies. 

\begin{figure}[h!]
\includegraphics[width=8.6cm]{ 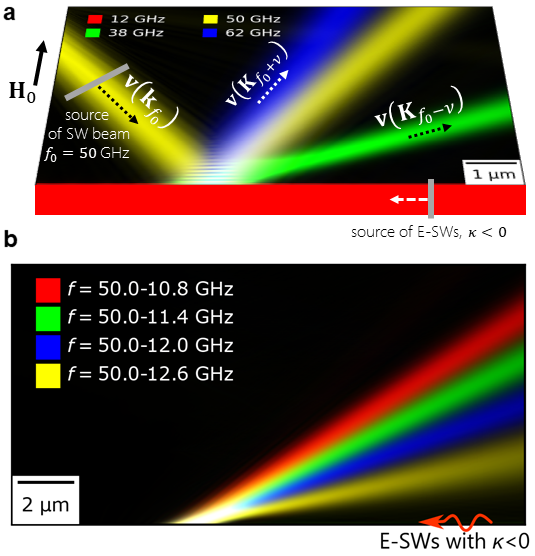}
\caption{
(a) The result of micromagnetic simulations showing the scattering of a 50~GHz B-SW beam (yellow color) incident at the angle of 30 degrees on a 12~GHz E-SW with $\kappa<0$ (red color). Two scattered beams at 38~GHz (green) and 62~GHz (blue) propagate at significantly different angles to each other, and to the reflected beam.
Areas, where excitation of the B-SW beam and E-SWs takes place, are schematically marked with gray lines.
(b) SW beams at frequencies 39.2~GHz (red color), 38.6~GHz (green color), 38~GHz (blue color), and 37.4~GHz (yellow color) excited at the edge of ferromagnetic film as the result of SW beam (at 50~GHz) scattering at the E-SWs of frequencies 10.8~GHz, 11.4~GHz, 12.0~GHz, and 12.6~GHz, respectively. 
% The figure is prepared by the analogy to the visible light, a white color denotes areas where at least two frequencies are mixed. The red arrow represents E-SW propagating leftwards.
}\label{fig:demux} 
\end{figure}

Interestingly, a slight modification of the previously studied system is enough to achieve such functionality, that is, to change the placement of the E-SWs source to the right side of the incident SW beam spot, see Fig.~\ref{fig:demux}(a).
In this case, the wavevector associated with the E-SWs has the negative sign ($\kappa<0$). Therefore, we expect the following pair of scattered beams: at frequency $f_0-\nu$ with $K_x=k_x+|\kappa|$ and at $f_0+\nu$ with $K_x= k_x-|\kappa|$. 
Isofrequency contour analysis shows that at $f_0-\nu$ the direction of group velocity may significantly  depend on the frequency of the E-SW $\nu$. However, due to the requirement to preserve the tangential to the film edge component of the wavevector and the finite value of the frequency at $k=0$, this process can only occur in a certain frequency range of E-SWs, i.e., for waves at frequencies greater than 10.8~GHz (the resonance frequency at $\kappa=0$) and lower than 13~GHz, at which the condition $\mathrm{max}\left\{ k_x(f-\nu)\right\} < k_x+|\kappa|$, still holds.
%\konstantininline{I think that if vec(K) is the wave vector of the scattered SW, then we need to change places of $k_x$ and $K_x$ in the paragraph above.}

Taking these limitations into account, we perform simulations for a system in which SW beam at 50~GHz frequency (incidence at 30$^\circ$) is scattered on  E-SWs excited at 10.8~GHz, 11.4~GHz, 12.0~GHz, and 12.6~GHz. Results of simulations are displayed in Fig.~\ref{fig:demux}(b), where scattered SWs at different frequencies are depicted by different colors. Indeed, depending on the frequency of the E-SW, the scattered SW beams propagate at easily distinguishable angles. For instance, the SW beam at frequency 39.2~GHz (red color) propagates at an angle around 33$^\circ$ with respect to the film edge, whereas the SW beam at frequency 37.4~GHz (yellow color) propagates at an angle around 11$^\circ$. 
%For the E-SWs at frequency 12.0~GHz \konstantininline{may be, 13~GHz. 12~GHz case is shown by blue color in Fig. 4.} (not shown here), the scattered SWs propagate along the interface. 
It means that E-SWs in a narrow 1.8~GHz  range of frequencies (from 10.8 up to 12.6~GHz) excite SW beams at frequencies from 37.4 up to 39.2~GHz that propagate under angles differing by as much as 30$^\circ$. 
Such functionality may be important for SW demultiplexing and frequency converting systems of future magnonic signal processing devices.

\subsubsection{Efficiency of the inelastic scattering process}
\begin{figure}[h!]
\includegraphics[width=8.6cm]{ 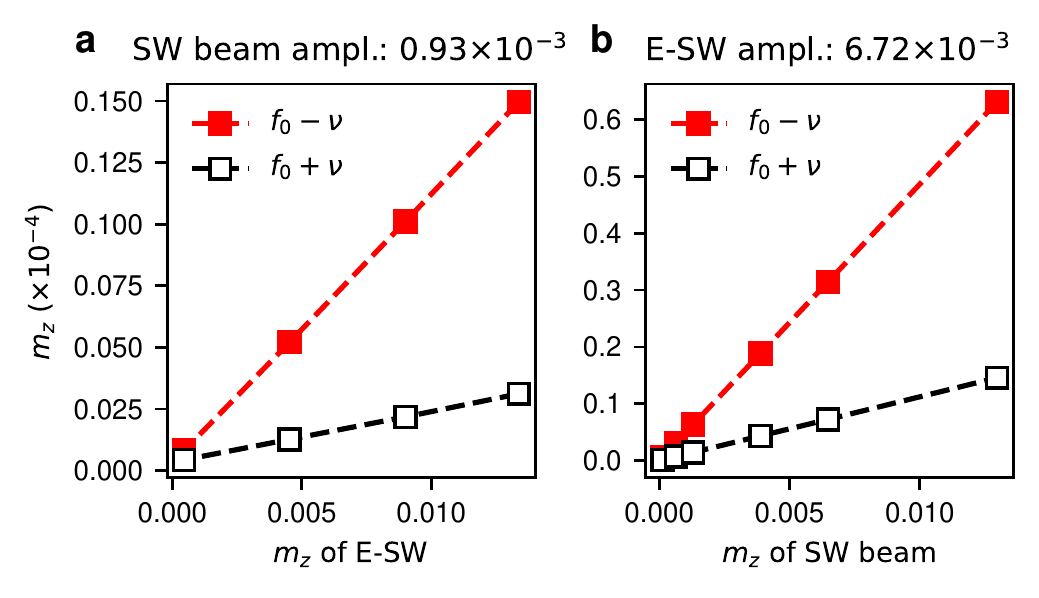}
\caption{The amplitudes of the $m_z$ component of magnetization of the inelastically scattered SWs at frequencies  $f_0\pm\nu$ for different amplitudes of $m_z$ of the E-SW (a) and the B-SW beam (b). The results of simulations presented in (a) are obtained for the B-SW beam of amplitude equal to $0.93\times 10^{-3}$, whereas the results presented in (b) are for the E-SW of the amplitude $6.72\times 10^{-3}$.
}\label{fig:Efficiency_E-SWs} 
\end{figure}

Let us discuss now the efficiency of the inelastic scattering process and compare qualitatively the simulation results with the analytical predictions. 
%In the following paragraphs we examine how the amplitude of scattered beams  depends on the amplitude of both the SW beam and E-SW. We also show how the scattering performance changes with the angle of incident SW beam.
According to the analytical model of the three-magnon scattering process developed in Sec.~\ref{Sec:model}, the amplitude of the scattered SW beam should linearly depend on the amplitude of both, the E-SW and the incident B-SW beam. In order to verify that, a set of micromagnetic simulations at different excitation amplitudes of the B-SW beam at frequency $f_0=50$~GHz and E-SWs at frequency $\nu=12$~GHz, and $\kappa>0$  is performed. Subsequently, the resulting amplitudes of the out-of-plane magnetization component, $m_z$,  at the distance $y=100$ nm from the film edge are analysed.  

The SW amplitude of the two  scattered beams at the frequencies $f_0-\nu$ and  $f_0+\nu$ on the amplitude of E-SWs are presented in Fig.~\ref{fig:Efficiency_E-SWs}(a). These results are obtained for the scattering of B-SW beam of amplitude maximum equal to $0.93\times 10^{-3}$ at the excitation point. As predicted by the model, the resulting linear relationship is clearly visible.
Further simulations also confirm  the linear dependence of the amplitude of the scattered B-SW beam on the amplitude of the incident SW beam, see Fig.~\ref{fig:Efficiency_E-SWs}(b). These results are obtained  at the E-SWs of amplitude equal to $6.72\times 10^{-3}$.   Interestingly, in both cases, the amplitude of SW beams at the frequency $f_0-\nu$ (the splitting process) is greater and faster increases with the amplitude of E-SW, than the amplitude of SW beam at the frequency  $f_0+\nu$ (the confluence process). Moreover, the amplitudes of the scattered beams are only 2-3 orders smaller than the amplitudes of the incident SW beam or E-SW. Clearly, the three-magnon inelastic scattering process under investigation is a thresholdless one.

\begin{figure}
\includegraphics[width=8.6cm]{ 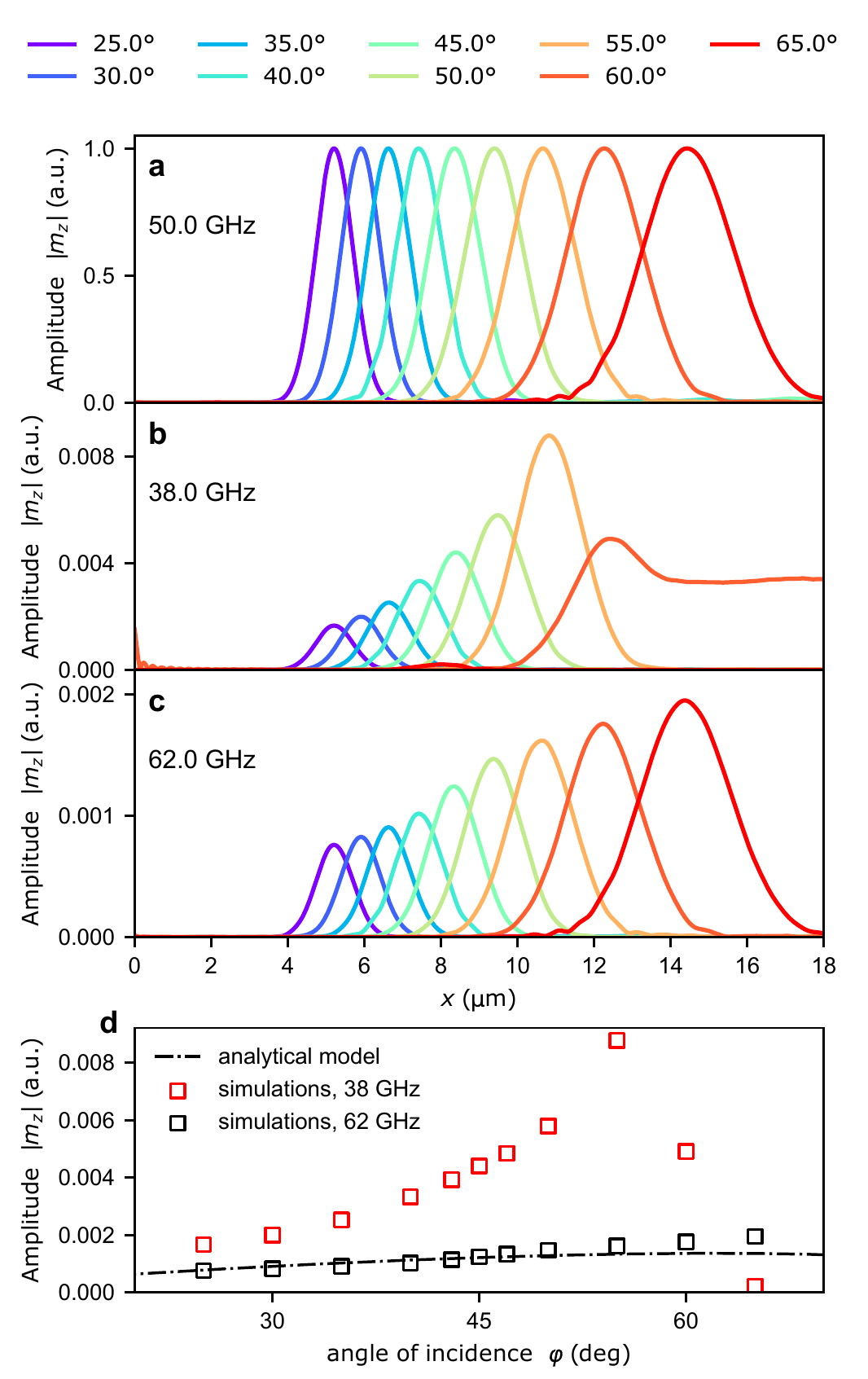}
\caption{\label{fig:fig_angleDep} 
(a), (b), and (c) dependencies of the $|m_z|$ component of dynamic magnetization detected at $y=100$ nm distance from the film edge for the reflection and inelastically scattered SWs at 50~GHz, 38~GHz, and 62~GHz, respectively. Different colours represent results at different angles of incidence of SW beam (50~GHz) falling at the edge, where E-SWs at frequency 12~GHz propagate along the $+x$ direction ($\kappa>0$). The amplitudes are normalized to the amplitude of the  elastically reflected beams at a given angle, i.e., $m_{\mathrm{scattered}}(y=100 \mathrm{~nm})/m_{\mathrm{reflected}}(y=100 \mathrm{~nm})$ where $m_{\mathrm{scattered}}(y=100 \mathrm{~nm})$ and $m_{\mathrm{reflected}}(y=100 \mathrm{~nm})$ are amplitudes of the scattered and reflected waves at $y=100$. (d) The comparison of the results of micromagnetic simulations (the empty black and red squares) with the analytical model (dash dotted black line) for inelastically scattered waves is done using the function
$|m_z| = C \mathrm{sin}(2\varphi) \chi((k_{y}-K_{y})\Delta )$
with $\Delta=L=10$ nm and $C=1.4\times 10^{-3}$.
}
\end{figure}

\begin{figure}
\includegraphics[width=8.6cm]{ 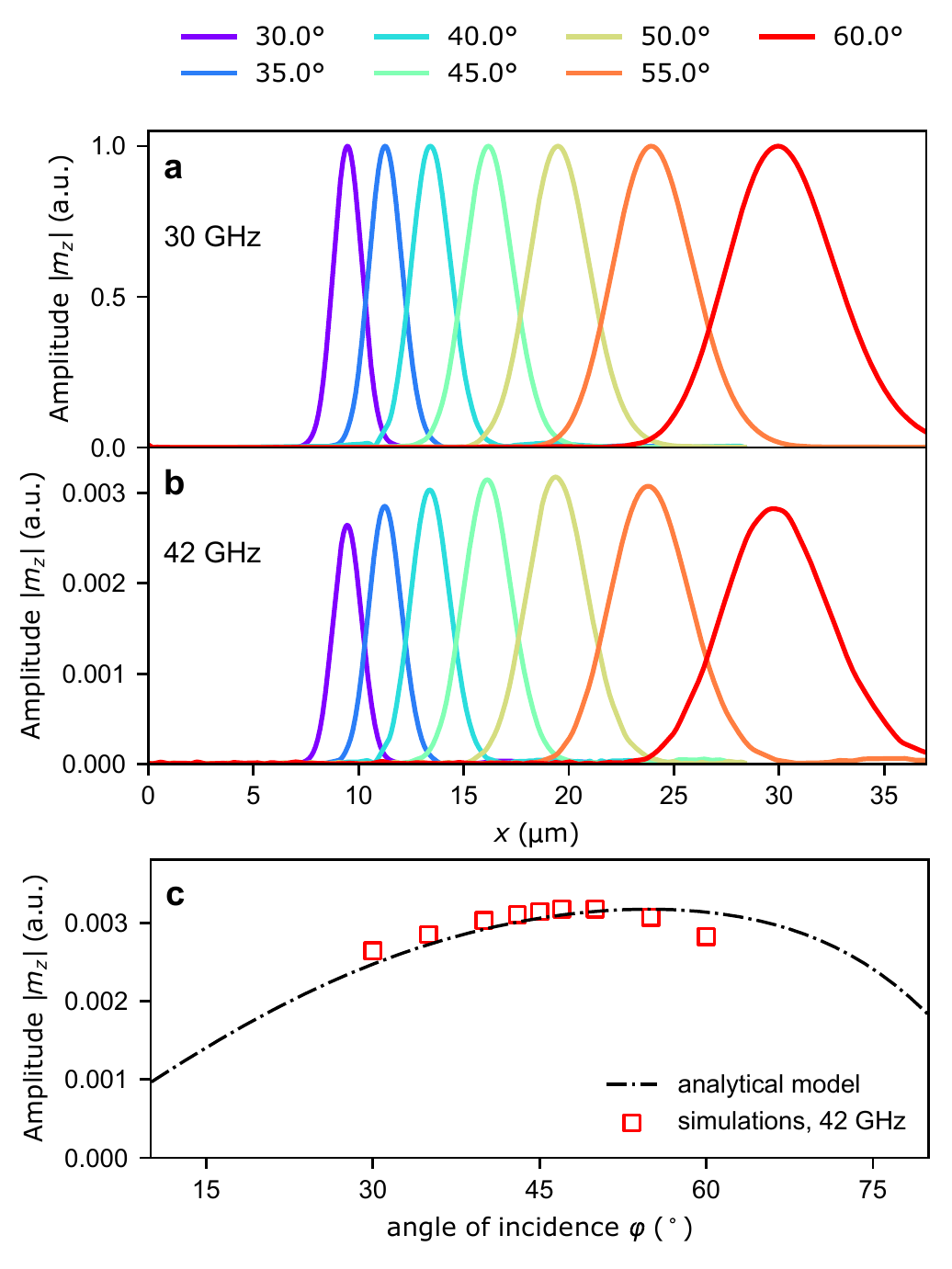}
\caption{\label{fig:fig_angleDep_30GHz} 
The dependencies of $|m_z|$ at $y=100$ nm distance from the film edge at the specular reflection [(a) 30~GHz] and inelastic scattering SW [(b) 42~GHz] propagating outwards the edge. Different colours represent results at different angles of the SW beam incidence at the edge where E-SW at frequency 12~GHz propagate. Similarly like in Fig.~\ref{fig:fig_angleDep}, the amplitudes are normalized to the amplitude of the  elastically reflected beams at a given angle. (c) The comparison of the results of micromagnetic simulations (the empty black squares)  with the analytical model (dash dotted black line) for inelastically scattered waves using the equation 
$|m_z| = C \mathrm{sin}(2\varphi) \chi((k_{y}-K_{y})\Delta )$
with $\Delta=L=10$ nm and $C=3.2\times 10^{-3}$.
}
\end{figure}

% \pawelinline{I am not sure if we need to include Fig.6(a)-(c) and Fig.7(a)-(b) in the manuscript.}
% \konstantininline{I agree. Put these figures to Suppl. Materials.}

As the next step, let us examine the amplitude of the inelastically scattered B-SW beams in dependence on the angle of incidence. Here, we consider the scattering  of B-SW beam at 50~GHz and 30~GHz on the E-SWs (12~GHz) of the amplitude  $6.72\times 10^{-3}$ [the same as used in Fig.~\ref{fig:Efficiency_E-SWs}(b)].

In Fig.~\ref{fig:fig_angleDep}(a), (b), and (c) we show the SW-beam cross-section $|m_z(x)|$ along the $x$-axis at $y=100$ nm for different angles of incidence (the different colors of lines) at 50~GHz -- specular reflection, 38~GHz, and 62~GHz, respectively. Since, the amplitude of the scattered beams in the studied range of amplitudes linearly depends on the amplitude of both E-SW and the incident SW beam, the amplitudes for each simulation are normalized to the SW amplitude of the elastically reflected beam. Thus, in Fig.~\ref{fig:fig_angleDep}(a), the amplitudes of all beams are equal 1, whereas, the amplitudes in Fig.~\ref{fig:fig_angleDep}(b), and (c) represent the relative value (see caption of the figure). It is visible, that with increasing the angle of incidence, the amplitudes of the scattered beams increase. Noteworthy, in the case of the scattered wave at the frequency $f_0 - \nu = 38$~GHz and for the angles of incidence greater than 55$^\circ$ the scattered SW beam is not observed. This is because at frequency 38~GHz there are no available solutions for the SW with a  high value of $K_x$, i.e., at $f=f_0-\nu$, $\mathrm{max}\{K\}<k_x-\kappa$, similarly to the case known in optics at total internal reflection there is no solution for the refracted waves.

At a lower frequency of the incident SW, i.e., 30~GHz, there are solutions only for $f_0+\nu=42$~GHz. The frequency $f_0-\nu=18$~GHz is only a bit larger than the bottom of the SW spectrum for B-SWs (16.5~GHz). Therefore, at the frequency 18~GHz only SWs with very small values of $K_x$, i.e., $K_x\ll k_x-\kappa$, can propagate.  The results are shown in Fig.~\ref{fig:fig_angleDep_30GHz}(a), and (b). The amplitude of the scattered beam reaches maximum for the angle of incidence equal 50$^\circ$ what is different result than in the case of 50~GHz SW beam.

The obtained angular dependence can be directly compared with the data from the analytical model. Combining Eq.~(\ref{eq:scatteringAmplitude_11}) with the dispersion relation, Eq.~(\ref{eq:Dispersion}), one may numerically derive scattering efficiency $B_{\mathbf{k}\boldsymbol{\kappa}\mathbf{K}}$ in dependence on the incidence angle of the B-SW beam. The comparison for the SW beams at frequencies 50~GHz and 30~GHz are presented in  Figs.~\ref{fig:fig_angleDep}(d) and \ref{fig:fig_angleDep_30GHz}(c), respectively. A  good agreement for both cases for the confluence process ($f_0+\nu$)  is achieved. 
Moreover, according to the analytical model, the maximum of the scattering efficiency should shift towards the waves incident at the angle of 45 degrees for very well localized waves and for even larger angles for poorer localization. This prediction also agrees with the simulation results. 

According to the analytical model, keeping $\Delta=10$~nm (see Eq.~(\ref{Eq:localization_condition})), the larger $k$ the larger $P(kL)$, and the smaller $|k_y-K_y|$ the larger $\chi((k_y-K_y)\Delta)$ ($|k_y-K_y|$ decreases with the increase of $f_0$ assuming constant $\nu$). Therefore, according to the analytical model, the larger efficiency of the confluence process should be observed for $f_0=50$~GHz. However, in simulations, the amplitude of the scattered waves in the confluence process is larger for 30~GHz waves. This is because the analytical model does not take into account the finite size of the beams  and the resulting differences in the areas where the scattering occurs between various frequencies. Indeed, the simulation results show that the  wider beam-spot width  at the film edge is for SWs at frequency 30 GHz than at 50 GHz. 
%This effect may explain this discrepancy between the simulation and analytical model results.
% This agrees well with the simulation results, since the wavelength for 30~GHz is greater than that for 50~GHz (33~nm for 50~GHz and 53~nm for 30~GHz), and therefore, for lower frequency, we get better localization. 

For the splitting process  ($f_0-\nu$) present at the 50~GHz incident SW beam, the simulated and calculated angular dependencies differ essentially. 
Overall, this means that although the analytical model describes the confluence process very well, it does not describe correctly the splitting process $(f_0 - \nu)$ obtained in micromagnetic simulations. This is not surprising since the analytical model does not take into account the presence of E-SW triggering the process of three-SW splitting.

\section{Conclusions}\label{Sec:conclusions}

We conducted a combined analytical and numerical study of the inelastic scattering of SW beam on the E-SW mode in the in-plane magnetized semi-infinite ferromagnetic film. We show that the scattering may lead to the appearance of the multiple SW beams at shifted frequencies related to the absorption and emission processes of the E-SW mode.

We presented the analytical model describing the three-magnon scattering process, which allows us to formulate the essential criteria of the E-SW mode absorption and emission, i.e., the energy and momentum conservation laws, and estimate the efficiency of the processes.  Subsequently, we confirmed analytical predictions by micromagnetic simulations demonstrating the scattering of 50~GHz SW beam on the E-SWs at frequency 12~GHz in 10 nm thick Py film. We showed that the observed inelastic scattering is a  three-magnon confluence or assisted (stimulated) splitting of the E-SW modes. Thus, the effect of the bulk-SW inelastic scattering on the E-SWs can be treated as a  magnonic analog of the Brillouin  scattering of electromagnetic waves by SWs.  Moreover, we found that the stimulated splitting process has considerably greater efficiency than the confluence process.

Importantly, depending on the chosen conditions, the inelastically scattered beams can propagate at the same or different angles in comparison with the elastically reflected SW beam. Based on the analysis of the isofrequency contours of SWs in the ferromagnetic film, we utilized the effect to demonstrate demultiplexing a signal carried by the E-SWs by separating spatially scattered SW beams at different frequencies. In contrast, the SW beam scattering by the standing E-SW can be used to split the signal into two beams.

The simplicity of the proposed system geometry and the linear dependence of the amplitude of the scattered SWs on the amplitudes of the E-SWs and the incident SW beam make the experimental realization feasible and promising for magnonic applications. The magnetostatic interactions being ubiquitous in ferromagnetic films and influencing SW dynamics to suggest the existence of the three-magnon processes between B-SW and other types of edge localized SWs.  Thus, our findings are important contributions to understanding and utilizing the nonlinear phenomena in SW dynamics and the next step towards the realization of magnonic circuits.

%designing of space-time metasurfaces\cite{hadad2015space,guo2019nonreciprocal,wang2020theory} operating on SWs. 

%\pawelinline{to discuss} \konstantininline{I suggest to put this discussion in the end of Sec. III.}
%Moreover, we found a very good agreement between the analytical model and micromagnetic simulations for scattering efficiency dependencies on the angle of incidence in the case of incident SWs at frequency 30~GHz, whereas the agreement is worse in the case of 50~GHz SWs. We consider that this disagreement for SWs with the larger frequency as a result of wavelength shortening... 

%\pawelinline{Perhaps we should extend it}

\begin{acknowledgments}
The research leading to these results has received funding from the National Science Centre of Poland, project no. 2019/35/D/ST3/03729. 
I.L. acknowledges support by COST action under project CA17123 MAGNETOFON.
K.G. acknowledges support by IKERBASQUE (the Basque Foundation for Science). The work of K.G. was supported in part by the Spanish Ministerio de Ciencia e Innovacion grant PID2019-108075RB-C33/AEI/10.13039/501100011033.
The simulations were partially performed at the Poznan Supercomputing and Networking Center (Grant No. 398).
\end{acknowledgments}
% \clearpage

\subsection*{DATA AVAILABILITY}
The data that support the findings of this study are available from the corresponding author
upon reasonable request.

\appendix
\section{Micromagnetic simulations}\label{Sec:AppSimulations}

Micromagnetic simulations have been proven to be an efficient tool for the calculation of magnetization dynamics in ferromagnetic materials. Presented results were obtained using the code MuMax3\cite{vansteenkiste2014design}, which solves time-dependent LLG Eq.~(\ref{eq:LLE}) with included Landau-Lifshits damping term, by using  the finite difference method. In simulations,
we considered an oblique SW beam propagation in Py thin film saturated by an in-plane magnetic field of value $\mu_0 H_0=0.3$ T oriented perpendicular to the film edge (see, Fig.~\ref{fig1_schema}). We assumed the typical magnetic parameters of Py (Ni$_{80}$Fe$_{20}$ alloy), i.e., the exchange stiffness constant $A=13$ pJ/m, saturation magnetization $M_{\mathrm{s}}=800$ kA/m, gyromagnetic ratio $|\gamma|=176$ rad~GHz/T and the value of damping $\alpha=0.0001$. The system of size $L_{x}\times L_{y}\times L_{z}$ is discretized with cuboid elements of dimensions $l_{x}\times l_{y}\times L_{z}$.
Lateral dimensions of the single cell $l_{x}\times l_{y}=5\times5$ nm$^2$ are slightly smaller than  the exchange length of Py ($l_\mathrm{ex}=5.7$ nm). The lateral dimensions of the system $L_x$ and $L_y$ are adjusted in such a way that the SW beams were clearly visible, e.g., in the case of 50~GHz SW, the beam incidences under angle of 30$^\circ$ what is the most frequently used scenario, $L_x=20$ \textmu m and $L_y=10$ \textmu m.

In the study, we simulated an SW beam of width ca. 1 \textmu m  in its waist propagating under the angle of incidence 50$^\circ$ (the angle between the SW wave-vector $\mathbf{k}_\mathrm{i}$ and the $y$-axis). SW beams were excited by the microwave magnetic field of the Gaussian profile type described in Ref.~[\onlinecite{gruszecki2014goos}]:
$h_\mathrm{mf}(x', y') = h_0 \mathrm{Win}(w x') \mathrm{exp}\left[ -\frac{ \left(y'/(0.5 L)\right)^2}{2\sigma^2} \right]$
where $h_0$ is the amplitude of the field, $\mathrm{Win}$ is the Dirichlet window function, $w=10$ nm is the width of the antennae, $L=2500$ nm is its length, and $\sigma^2=0.1$.

The point-source of radius ca. 30 nm exciting E-SWs was located at the edge of the film  at the left or/and right side of the beam spot in distance ca. 5 \textmu m from the incident beam spot. At the borders of the simulated domain, different than the film's edge, the absorbing boundary conditions have been assumed\cite{gruszecki2018mirage}.

The simulations consisted of three stages. In the first one a static magnetization configuration was obtained, then, the SWs (B-SW beam and E-SW) were continuously excited until reaching a steady-state (typically we run simulations for 60 ns). Finally, the simulation results ($m_z$ component of the magnetization) were stored with the sampling interval of 5 ps to the matrix: $m_z\left(t;x,y\right)$. The results obtained in such a way were next processed. 

\textbf{Postprocessing.} 
In the first stage of postprocessing we calculated the pointwaise fast Fourier transforms (FFT) over time of $m_z$ component of magnetization sampled: $\widetilde{m}_z(f;x,y)=\frac{2}{N}\mathcal{F}_t \left[ m_z(t; x, y) \right] $, where $\mathcal{F}_t$ denotes FFT  operation over time implemented in NumPy\cite{harris2020array, numpyFFT} and  $N$ is the number of samples processed. 
We have chosen the following sampling scheme providing frequency resolution $\delta f=0.2$~GHz, i.e., we have used $N=1200$ samples stored with sampling interval $t_\mathrm{sampl}=1/(N \delta f)$. 
The normalization $\frac{2}{N}$ together with the sampling scheme precisely resolving frequencies used in the study enabled us to precisely resolve the amplitudes of SWs after FFT. Namely, there is the same amplitude of the time-dependent signal of a given frequency as the absolute value of the peak corresponding to that frequency of the transformed signal.
In all the further steps this matrix was processed.

\textbf{SW spectrum.}
The SW spectra presented in Fig.~\ref{fig2_beams} was obtained using the following formula $|m_z|(f) = \mathrm{max}(| \widetilde{m}_z(x,y)|)(f)$. The amplitude corresponds to maximal amplitudes of SW at given frequency $f$ present in the sample.

\textbf{Mode profiles.} Mode profiles for frequencies $f_i$ presented in Fig.~\ref{fig2_beams}(b)-(d) were obtained using the formula $m_{z,f_i} = \left| \widetilde{m}_z(f=f_i; x,y) \right|$.

\textbf{Mode profiles in the $k$-space.} The isofrequency map (mode profile in the $k$ space) for frequency $f_i$ can derived from the SW mode profiles as a 2-dimensional-FFT: 
$IFM_{f_i}(k_x, k_y)= \left| \mathcal{F}_{x,y} \left[ \mathrm{Re}\Big\{ \widetilde{m}_z(f=f_i;x,y) \Big\} \right] \right| $, where $\mathcal{F}_{x,y}$ denotes 2-dimensional FFT operation over the $x$ and $y$ coordinates.

\textbf{Dispersion relation.} To simulate the dispersion relation presented in Fig.~\ref{fig1_schema}(b), the SW dynamics were excited using a broadband low-amplitude microwave field of gaussian-like spatial profile located just at the edge (within potential well created by the demagnetizing field) with a half-width at half maximum of about 10 nm. The time dependence of the microwave field was as follows: $h_\mathrm{mf} \propto \mathrm{sinc}(2 \pi f_\mathrm{cut} (t-t_0))$, where $f_\mathrm{cut}=70$~GHz is the cut-off frequency and $t_0=10/f$ is the time delay. The results of these simulations (namely $m_z$-component of magnetization) were stored at a sampling frequency of $f_\mathrm{sampl}=150$~GHz. 

Then, for each $y$-coordinate, the 2-dimensional-FFT over the $x$-coordinate and time $t$ was calculated.  This allowed the determination of the three-dimensional complex
matrix  $\tilde{m_z}(y; f, k_x)$ depending on the $y$-coordinate, frequency $f$, and wavevector component $k_x$. The final step was to calculate the mean value over the $y$-coordinate from the absolute value of the matrix $\tilde{m_z}$: $D(f, k_x)=\left<|\tilde{m_z}(y; f, k_x)|\right>_y$.  The resulting $D(f, k_x)$ matrix consists of positive real numbers and represents the dispersion relation.

\bibliography{literature}% Produces the bibliography via BibTeX.

\end{document}